A generalized lattice Boltzmann model for flow through tight porous media with Klinkenberg's effect

Li Chen [a,b], Wenzhen Fang [a], Qinjun Kang [b], Jeffrey De'Haven Hyman [b], Hari S Viswanathan [b], Wen-Quan Tao [a]

a: Key Laboratory of Thermo-Fluid Science and Engineering of MOE, School of Energy and Power Engineering, Xi'an Jiaotong University, Xi'an, Shaanxi, 710049, China b: Computational Earth Science, EES-16, Earth and Environmental Sciences Division, Los Alamos National Laboratory, Los Alamos, New Mexico, 87544, USA

**Abstract:** Gas slippage occurs when the mean free path of the gas molecules is in the order of the characteristic pore size of a porous medium. This phenomenon leads to the Klinkenberg's effect where the measured permeability of a gas (apparent permeability) is higher than that of the liquid (intrinsic permeability). A generalized lattice Boltzmann model is proposed for flow through porous media that includes Klinkenberg's effect, which is based on the model of Guo et al. (Z.L. Guo et al., Phys.Rev.E 65, 046308 (2002)). The second-order Beskok and Karniadakis-Civan's correlation (A. Beskok and G. Karniadakis, Microscale Thermophysical Engineering 3, 43-47 (1999), F. Civan, Transp Porous Med 82, 375-384 (2010)) is adopted to calculate the apparent permeability based on intrinsic permeability and Knudsen number. Fluid flow between two parallel plates filled with porous media is simulated to validate model. Simulations performed in a heterogeneous porous medium with components of different porosity and permeability indicate that the Klinkenberg's effect plays significant role on fluid flow in low-permeability porous media, and it is more pronounced as the Knudsen number increases. Fluid flow in a shale matrix with and without fractures is also studied, and it is found that the fractures greatly enhance the fluid flow and the Klinkenberg's effect leads to higher global permeability of the shale matrix.

# I. INTRODUCTION

Fluid flow and transport processes in porous media are relevant in a wide range of fields, including hydrocarbon recovery, ground water flow, $CO_2$ sequestration, metal foam, fuel cell, and other engineering applications [1]. Understanding the fluid dynamics in porous media and predicting effective transport properties (permeability, effective diffusivity, etc) is of paramount importance for practical applications.

Fluid flow and transport in porous media are usually observed physically and treated theoretically at two different scales: pore scale and representative elementary volume (REV) scale [2-6]. A REV of a porous medium is the smallest volume for which large fluctuations of observed quantities (such as porosity and permeability) no longer occur and thus scale characteristics of a porous flow hold [4]. Multiple techniques are in use for numerically modeling fluid flow in porous media at different scales. Conventionally, fluid flow through porous media is solved by discrete numerical methods (such as finite difference, finite volume and finite element methods) based on governing partial differential equations such as Navier-Stokes (or Stokes) equation, Darcy equation and extended Darcy equations (Brinkman-Darcy and Forchheimer-Darcy equations). The lattice Boltzmann method (LBM) is an alternative and efficient tool for simulating such processes. It has shown enormous strengths over conventional numerical methods to study complicated fluid flow such as in complex structures and multiphase flow [7-9]. The LBM has been widely applied to simulate fluid dynamics in porous media at the pore scale where the LB equation recovers the Navier-Stokes equation and solid matrix is usually impermeable [9-18]. However, the pore-scale LB model is unrealistic to perform REV scale simulation of porous medium flow due to the huge computational resources required. Therefore, recently several REV-scale LBM models have been proposed [2-5], which enhance the capacity of LBM for larger scale applications. These models recover the common continuum equations for fluid flow in porous media such as Darcy equation and extended Darcy equations [2-5]. Usually, force schemes are adopted in LB to account for the presence of the porous media. Using the force scheme proposed in [19], Guo et al. developed a generalized LB model for fluid flow through porous media, where the generalized Navier-Stokes equation proposed in [20], including the Brinkman term, the linear (Darcy) and nonlinear (Forchheimer) drag terms, can be recovered in the incompressible limit [4].

Permeability $k$ is a key variable to describe the transport capacity of a porous medium, and is required in the REV scale simulations [20]. The intrinsic permeability of a porous medium, for which there is no slip on the fluid-solid boundary, only depends on the porous structures. However, when Knudsen number ($Kn$, ratio between the mean free path of gas and the characteristic pore size of a porous medium) is relatively high, the gas molecules tend to slip on the solid surface. Gas slippage in porous media and its effects on permeability was first studied by Klinkenberg [21]. It was found that due to the slippage phenomenon, the measured gas permeability (apparent permeability) through a porous medium is higher than that of the liquid (usually called intrinsic permeability), and the difference becomes increasingly important as $Kn$ increases. This phenomenon is called Klinkenberg's effect. Klinkenberg proposed a linear correlation between apparent permeability and the reciprocal of the pressure [21]. This correlation has been a consistent basis for the development of new correlations between the apparent and intrinsic permeability [22-25]. Later, Karniadakis and Beskok [26] developed a second-order correlation based on fluid flow in micro-tubes, which was shown to be valid over the flow regimes that include: Darcy flow ($Kn<0.01$) regime, slip flow regime ($0.01<Kn<0.1$), transition flow regime ($0.1<Kn<10$) and free molecular flow regime ($Kn>10$). Recent experimental studies of natural gas through tight porous rocks found that the apparent permeability can be one hundred times higher than the intrinsic permeability, emphasizing the importance of gas slippage in the study of the fluid flow in tight porous rocks [27]. Therefore, if Klinkenberg's effect is neglected, the transport rate of gas in tight rocks will be greatly underestimated. However, none of the existing REV-scale LB models account for the Klinkenberg's effect.

The present work is motivated by the multi-scale fluid flow in a shale matrix. Gas-bearing shale formations have become major sources of natural gas production in North America, and are expected to play increasingly important roles in Europe and Asia in the near future [28]. Experimental observations indicate that the shale matrix is composed of pores, non-organic minerals (predominantly clay minerals, quartz, pyrite) and organic matter [29-33]. Different components have different structural and transport properties. While the non-organic matter is usually impermeable, nanosize pores widely exist in the organic matter, with pore diameters in the range of a few nanometers to hundreds of nanometers [30], thus allowing transport to occur through the bulk shale matrix. These pores are so small that gas slippage occurs therein [28, 34, 35]. Therefore, apparent permeability, rather than intrinsic permeability, should be defined and used in the REV scale studies of shale gas transport [28]. Therefore, the accurate prediction of shale matrix permeability is crucial for improving the gas production and lowering production cost.

In the present work, a generalized LB model for fluid flow through porous media with Klinkenberg's effect is developed based on the work of Guo et al. [4]. The following aspects of flow in porous media are investigated: how does the permeable matrix impact the fluid flow in the porous medium? How does Klinkenberg's effect influence the flow filed and the apparent permeability of the porous medium? Why fractures are important for enhancing permeability? The remaining parts of this study are arranged as follows. The generalized Navier-Stokes equations that include Klinkenberg's effect are developed in Section II. The generalized LB model for solving the generalized Navier-Stokes equations is introduced in Section III. In Section IV, first the model is validated by simulating fluid flow between two parallel plates filled with a porous medium. Then, a porous medium with three different components is reconstructed, and fluid flow therein is investigated. The flow filed and influence of Klinkenberg's effect are discussed in detail. Finally in this section, the importance of fractures on fluid flow in porous media is also illustrated. Finally, some conclusions are drawn in Section V.

## II. GENERALIZED MODEL FOR POROUS FLOW WITH KLINKENBERG'S EFFECT

### A. Generalized Navier-Stokes equations

The generalized Navier-Stokes equations proposed by Nihiarasu et al. [20] for isothermal incompressible fluid flow in porous media can be expressed as follows

$$\nabla \cdot \mathbf{u} = 0 , \qquad (1a)$$

$$\frac{\partial \mathbf{u}}{\partial t} + (\mathbf{u} \cdot \nabla)\frac{\mathbf{u}}{\varepsilon} = -\frac{1}{\rho}\nabla(\varepsilon p) + \upsilon_e \nabla^2 \mathbf{u} + \mathbf{F}, \tag{1b}$$

where $t$ is time, $\rho$ volume averaged fluid density, $p$ volume averaged pressure, $\mathbf{u}$ the superficial velocity, $\varepsilon$ the porosity and $\upsilon_e$ an effective viscosity equal to the shear viscosity of fluid $\upsilon$ times the viscosity ratio $J$ ($\upsilon_e = \upsilon J$). The force term $\mathbf{F}$ represents the total body force due to the presence of the porous media and other external body forces

$$\mathbf{F} = -\frac{\varepsilon \upsilon}{k}\mathbf{u} - \frac{\varepsilon F_\varepsilon}{\sqrt{k}}|\mathbf{u}|\mathbf{u} + \varepsilon \mathbf{G}, \tag{2}$$

where $\upsilon$ is fluid viscosity and $\mathbf{G}$ is the external force. The first term on RHS is the linear (Darcy) drag force and the second term is the nonlinear (Forchheimer) drag force. The geometric function $F_\varepsilon$ and the permeability $k$ are related to the porosity of the porous medium, and for a porous medium composed of solid spherical particles, they can be calculated by Ergun's equation [4]. The quadratic term (Forchheimer term) becomes important when the fluid flow is relatively strong and inertial effects become relevant.; it can be neglected for fluid flow with Reynold number much lower than unity.

### B. Klinkenberg's effect

Gas slippage is a phenomenon that occurs when the mean free path of a gas particle is comparable to the characteristic length of the domain. Klinkenberg [21] first conducted the study of gas slippage in porous media, and found that the permeability of gas (apparent permeability) through a tight porous medium is higher than that of liquid due to Gas slippage. Klinkenberg [21] proposed a linear correlation for correcting the gas permeability

$$k_a = k_d f_c, \tag{3}$$

where $k_d$ is called Klinkenberg's corrected permeability, which is the permeability of liquid, or the intrinsic permeability. $k_d$ only depends on the porous structures of a porous medium and is not affected by the operating condition and fluid properties. The correction factor $f_c$ is given by [21]

$$f_c = (1 + \frac{b_k}{p}), \tag{4}$$

where Klinkenberg's slippage factor $b_k$ depends on the molecular mean free path $\lambda$, characteristic pore size of a porous medium $r$, and pressure [21]

$$\frac{b_k}{p} = \frac{4c\lambda}{r} \approx 4Kn, c \approx 1, \tag{5}$$

It can be found in Eqs. (3-5) that the apparent permeability $k_a$ not only depends on the topology of a porous medium, but also is affected by pressure and temperature conditions. Based on Eq. (4), various expressions of $b_k$ have been proposed in the literature. Heid et al. [22] and Jones and Owens [23] proposed similar expressions by relating $b_k$ to $k_d$. Sampath and Keighin [24] and Florence et al. [25] considered the effects of porosity and developed a different form by relating $b_k$ to $k_d/\varepsilon$. Klinkenberg's correlation is a first-order correlation. Beskok and Karniadakis [26] developed a second-order correlation, which has been shown to be capable of describing the four fluid flow regimes (viscous flow ($Kn<0.01$), slip flow($0.01<Kn<0.1$), transition flow($0.1< Kn <10$), and free molecular flow($Kn >10$))

$$f_c = (1 + \alpha(Kn)Kn)\left[1 + \frac{4Kn}{1 - bKn}\right], \tag{6}$$

where slip coefficient $b$ equals -1 for slip flow, and $\alpha(Kn)$ is the rarefaction coefficient. The expression of $\alpha(Kn)$ is very complex in [26], and later Civan [36] proposed a much simplified one

$$\alpha(Kn) = \frac{1.358}{1+0.170Kn^{-0.4348}}, \tag{7}$$

We refer to the combination of Eq. (6) with Eq. (7) as the Beskok and Karniadakis-Civan's correlation.

## C. Generalized model for porous flow with Klinkenberg's effect

In the present study, Eq. (2) is modified to incorporate the effects of gas slippage phenomenon (Klinkenberg's effect) by substituting $k$ with apparent permeability $k_a$

$$\mathbf{F} = -\frac{\varepsilon \upsilon}{k_a}\mathbf{u} + \varepsilon \mathbf{G}. \tag{8}$$

Compared with Eq. (2), the nonlinear drag force is not considered in Eq. (8), because usually flow rate is extremely low in low-permeability tight porous media such as shale matrix [35]. We use the Kozeny-Carman (KC) equation [37] to calculate the intrinsic permeability of the porous medium

$$k_d = C\frac{\varepsilon^3}{(1-\varepsilon)^2}, \tag{9}$$

with $C$ is the KC constant and is set as $d^2/180$ for packed-spheres, where $d$ is the diameter of the solid spheres. It is worth mentioning that empirical equations developed for a specific porous medium can also be used to calculate the corresponding intrinsic permeability. After $k_d$ is determined, Beskok and Karniadakis-Civan's correlation Eqs. (6-7) is then used to calculate $k_a$. To calculate $Kn$ in Eqs. (6-7), the mean free path $\lambda$ and the characteristic pore radius of the porous medium $r$ should be determined. The former one is calculated by [38]

$$\lambda = \frac{\mu}{p}\sqrt{\frac{\pi RT}{2M}} \qquad (10)$$

where $R$ is the gas constant, $T$ the temperature and $M$ the molar mass. Following [27], the following expression proposed by Herd et al. is used to calculate $r$ [22]

$$r = 8.886\times 10^{-2}\sqrt{\frac{k_{\mathrm{d}}}{\varepsilon}} \qquad (11)$$

The units of $r$ and $k$ are μm and mD (1mD=9.869×10$^{-16}$ m$^2$), respectively. We have to admit that Eq. (11) is based on the assumption that pores in the porous media are uniform, parallel and cylindrical capillaries, and thus it only provides a approximation of the pore radius [22, 27]. More accurate expressions instead of Eq. (11) can be adopted to improve the prediction of $r$, which, although out of the scope of the present study, deserve further study.

On the whole, Eq. (1), combined with Eq. (8), together with apparent permeability calculated by Eqs. (3) and (6-7), is called the generalized Navier-Stokes equations for porous flow that includes Klinkenberg's effect.

### III. LATTICE BOLTZMANN MODEL

In this section, we develop our LB model for solving the above-proposed generalized Navier-Stokes equations for porous flow with Klinkenberg's effect based on the work of Guo et al. [4] because porosity is the major input for this model. The evolution of the density distribution function in the LB framework is as follows

$$f_i(\mathbf{x}+\mathbf{e}_i\Delta t, t+\Delta t) - f_i(\mathbf{x},t) = -\frac{1}{\tau}(f_i(\mathbf{x},t) - f_i^{\mathrm{eq}}(\mathbf{x},t)) + \Delta t F_i \qquad (12)$$

where $f_i(\mathbf{x},t)$ is the $i$th density distribution function at the lattice site $\mathbf{x}$ and time $t$. $\Delta t$ is lattice time step. The dimensionless relaxation time $\tau$ is related to the viscosity. The lattice discrete velocities $\mathbf{e}_i$ depend on the particular velocity model. For the **D2Q9** model with nine velocity directions at a given point in two-dimensional space, $\mathbf{e}_i$ are given by

$$\mathbf{e}_i = \begin{cases} 0 & i=0 \\ (\cos[\frac{(i-1)\pi}{2}], \sin[\frac{(i-1)\pi}{2}]) & i=1,2,3,4 \\ \sqrt{2}(\cos[\frac{(i-5)\pi}{2}+\frac{\pi}{4}], \sin[\frac{(i-5)\pi}{2}+\frac{\pi}{4}]) & i=5,6,7,8 \end{cases} \quad (13)$$

The equilibrium distribution functions $f^{eq}$ are of the following form by considering the effects of porosity [4]

$$f_i^{eq} = \omega_i \rho \left[1 + \frac{3}{c^2}(\mathbf{e}_i \cdot \mathbf{u}) + \frac{9}{2\varepsilon c^4}(\mathbf{e}_i \cdot \mathbf{u})^2 - \frac{3}{2\varepsilon c^2}\mathbf{u}^2\right] \quad (14)$$

where the weight factors $\omega_i$ are given by $\omega_0=4/9$, $\omega_{1-4}=1/9$, and $\omega_{5-8}=1/36$. Different schemes have been proposed to incorporate the force term (Eq. (8)) into the LB model, such as the modified equilibrium velocity method by Shan and Chen [39], the exact difference method (EDM) scheme [40] and Guo's force scheme [19]. Among them, Guo's force scheme can recover the exact Navier-Stokes equation without any additional term [19], which is thus adopted to calculate the force term in Eq. (12)

$$F_i = \omega_i \rho (1 - \frac{1}{2\tau}) \left[\frac{3}{c^2}(\mathbf{e}_i \cdot \mathbf{F}) + \frac{9}{\varepsilon c^4}(\mathbf{e}_i \cdot \mathbf{u})(\mathbf{e}_i \cdot \mathbf{F}) - \frac{3}{c^2}(\mathbf{u} \cdot \mathbf{F})\right] \quad (15)$$

Accordingly, the fluid velocity and density are defined as

$$\rho = \sum_i f_i \qquad (16a)$$

$$\rho \mathbf{u} = \sum_i f_i \mathbf{e}_i + \frac{\Delta t}{2} \rho \mathbf{F} \qquad (16b)$$

Note that **F** also contains the velocity, as can be seen in Eq. (8). Due to only linear drag term considered in Eq. (8), Eq. (16b) is a linear equation, and thus the velocity can be easily solved

$$\mathbf{u} = \frac{\sum_i f_i \mathbf{e}_i + \frac{\Delta t}{2} \rho \varepsilon \mathbf{G}}{\rho + \frac{\Delta t}{2} \frac{\varepsilon \upsilon}{k_a} \rho} \qquad (17)$$

If the nonlinear drag term is further considered, Eq. (16b) turns into a quadratic equation [4]. Eqs. (12), (14) and Eq. (15) can recover the generalized Navier-Stokes equation in Eq. (1) using Chapman–Enskog multiscale expansion under the low Mach number limit[41].

This LB model incorporates the influence of porous media by introducing a newly defined equilibrium distribution function (Eq. (14)) and adding a force term (Eq. (15)) into the evolution equation (Eq. (12)), and thus it is very close to the standard LB model [4]. Without invoking any boundary conditions, it can automatically model the interfaces between different components in a porous medium with spatially variable porosity and permeability [4]. In the simulation of flow through a porous system, this model is employed by simply replacing the usual computational nodes with porous medium nodes in the region occupied by the porous medium. Each node in this region is given a porosity based on the experimental results. Intrinsic permeability and apparent permeability are calculated based on the scheme introduced in Section II. At a node in pore space, the porosity is unity and the Darcy term is zero. Under such condition, Eq. (1b) reduces to the Navier-Stokes equation for free fluid flow. At a node in the impermeable component, the drag force is specified to be infinity, and the velocity is zero according to Eq. (1b). Such flexibility of the proposed model allows for it to automatically simulate interfaces between different components in a porous medium [4]. Compared to the original model of Guo et al. [4], the Klinkenberg's effect is taken into account by adopting the second-order Beskok and Karniadakis-Civan's correlation to calculate the apparent permeability, allowing the current model to handle fluid flow in porous media with gas slippage at the REV scale. When *Kn* is small, the apparent permeability is close to the intrinsic permeability, and Klinkenberg's effect can be neglected according to Eqs. (3). Under such circumstances, the model reduces to the original model in Ref. [4]. In summary, the proposed generalized model can be used to simulate at REV scale wide flow regimes by considering the Klinkenberg's effect.

## IV. RESULTS AND DISCUSSION
### A. Flow between two parallel plates filled with a porous medium

Fluid flow between two parallel plates filled with a porous medium of porosity $\varepsilon$ is simulated to validate the present LB model and to illustrate the Klinkenberg's effect. The flow is driven by pressure difference $\Delta p$ at the inlet and outlet. The flow at steady state is described by the following Brinkman-extended Darcy equation [4]

$$\frac{\upsilon_e}{\varepsilon}\frac{\partial^2 u}{\partial y^2} - \frac{\upsilon}{k_a}u + G = 0 \qquad (18)$$

with $u(x,0)=u(x,H)=0$, where $H$ is distance between the two plates and $G$ is the external body force. The velocity in the y direction is zero everywhere. The analytical solution for fluid flow under these conditions is

$$u = \frac{GK}{\upsilon}(1 - \frac{\cosh[a(y-H/2)]}{\cosh(aH/2)}), a = \sqrt{\frac{\upsilon\varepsilon}{\upsilon_e K}} \qquad (20)$$

where cosh is the hyperbolic function with $\cosh(x) = (e^x + e^{-x})/2$. The external body force can be calculated by the pressure difference according to $G=\Delta p/L/\rho$, where $L$ is the length of the plates. In all the simulations of the present study, the viscosity ratio $J$ is assumed to be unity, thus $v_e$ equals $v$.

Fig. 1(a) shows the characteristic pore radius calculated by Eq. (11) and $Kn$ under different porosity. $d$ in Eq. (9) for calculating the intrinsic permeability is set as 40nm. The pore radius decreases as the porosity is reduced, which is expected based on Eq. (9) and (11). For the minimum porosity studied (0.02), the pore radius is only about 1.12 nm (a typical order of the size of throats connecting larger pores in the organic matter of shale matrix [42]). To calculate the mean free path in Eq. (10), the temperature is set as 323 K, the molar mass $M$ is that of methane $16×10^{-3}$ kg mol$^{-1}$, and the viscosity $\mu$ of methane is determined by using a online software called Peace Software [43]. Three pressures, 4000psi, 1000psi and 100psi, are studied (1psi≈6894.75Pa). The mean free path under the three pressures is 0.38, 1.09 and 10.26nm, respectively. As shown in Fig. 1(a), $Kn$ increases nonlinearly as the porosity decreases. At a fixed value of porosity, $Kn$ is higher for lower pressures due to the longer mean free path. The flow can enter the transition regime (0.1<$Kn$<10) when the porosity is low. The correction factor increases as the porosity decreases (or $Kn$ increases) as shown in Fig. 1(b), which is expected based on the Beskok and Karniadakis-Civan's correlation Eq. (6). For the highest pressure of 4000psi, the value of the correction factor is about 2. This value can be as high as 60 when the pressure is reduced to 100psi.

Fig. 2 shows the numerical simulation results as well as the analytical solutions for the velocity profile between the two plates. In the simulations, the domain is discretized by 200×80 square meshes. The relaxation time $\tau$ in LB is set as 0.9. No-slip boundary conditions are used for the top and bottom walls and a pressure difference is applied between left inlet and right outlet. When porosity is 1.0, there is no porous medium and the fluid flow is free fluid flow between the two plates. Therefore, the velocity profile is the same for all the cases. When the porosity is reduced, the velocity profile flattens due to the presence of the porous medium. As the Klinkenrg's effect becomes significant with the decrease of the pressure, the boundary layer becomes thicker, as shown in Fig. 2. For all the cases, the simulation results (symbols) are in good agreement with the analytical solutions (lines), which confirms the validity of the present model.

Fig. 3 shows the relationship between the permeability of the entire domain (called global permeability in the present study, obtained by applying Darcy' law to the entire domain) and the porosity. This is an important graph clearly demonstrating the effects of Klinkenberg's effect on the domain scale. For fluid flow between two plates without porous media, the global permeability is $H^2/12$ according to the cubic law [5]; the simulated global permeability is 530.68, very close to the analytical solution obtained from the cubic law (533.3), further demonstrating the validity our model. The important observation from Fig. 3 is that the global permeability increases as the pressure decreases, due to higher $Kn$ and thus larger apparent permeability of the porous medium. Besides, the permeability difference between different cases increases as the porosity is reduced. For porosity of 0.2, two orders of magnitude of the permeability difference can be observed between the case with $p$=100 psi and that without Klinkenberg's effect. Therefore, apparent permeability, rather than the intrinsic permeability, must be adopted to simulate the fluid flow in porous media with gas slippage, especially under low pressure.

### B. Flow in porous media with different components

In this section, the flow through a trimodal heterogeneous porous medium is studied. Three components coexist in the porous medium including the pores (light blue), impermeable solids (black) and permeable solids (green), as shown in Fig. 4. The impermeable solids are generated using a self-developed overlapping tolerance circle method based on the three-dimensional overlapping tolerance sphere method [44]. The permeable solids are generated using the quartet structure generation set (QSGS) method [45]. This porous system roughly represents the structures of the shale matrix composed of pores, nonorganic matter and organic matters [29-33]. In shale matrix, the organic matter is the source of shale gas (methane) and plays an important role in shales. Fig. 4(b) shows the nanoscale structures of the organic matter reproduced from the literature (Fig. 9 in [33]), which is a magnified image of the organic matter in Fig. 4(a) (as schematically shown in the red circle of Fig. 4(a)). Numerous nanosize pores can be observed in Fig. 4(b), with pore diameters in the range from a few nanometers to hundreds of nanometers. In some shales almost all the pores in the shale matrix are associated with organic matter, and thus the permeability of the organic matter is very important for shale gas transport. The pore structures of these organic matters are geometrically and topologically intricate being the result of several factors including maturity, organic composition and late localized compaction [29, 30]. Even consistent values of porosity are elusive. Loucks et al. reported a range of porosity 0%~30% in the organic matter of Barnett shales, North Texas [30], while Sondergeld et al. estimated a porosity of 50% of the organic matter from Barnett shales [32]. In the present study, a wider range of porosity ($0.1 \leq \varepsilon \leq 0.9$) of the organic matter is studied. In Fig. 4, volume fraction of the impermeable solid and permeable solid is 0.4 and 0.3, respectively. There is no connected void space percolating the *x* direction. The domain size is 410*200 lattices with a resolution of each lattice as 10nm. With such a coarse resolution, nanosize pores in the organic matters cannot be fully resolved, and thus the REV scale model is adopted. Periodic boundary conditions are applied on the top and bottom walls. For a void space node the porosity is unit, while for a impermeable solid node the porosity is zero, leading to zero and infinite drag force, respectively, according to Eqs. (8-9). Other parameters and boundary conditions are the same as that in Section IV. A.

Fig. 5 shows the distributions of velocity magnitude |**u**| inside the porous medium without (a) and with increasingly stronger (from (b) to (d)) Klinkenberg's effect. The porosity of the permeable solid for the set of images in the left, middle and right column is 0.8, 0.5 and 0.1, respectively. When porosity of the permeable solid is relatively high ($\varepsilon_{ps}$=0.8 in the left column, subscript "ps" stands for "permeable solid"), the Darcy drag term in the permeable solid is small and the local flow resistance is weak. Therefore, fluid can flow through the permeable solids. In such cases, there are two main preferred pathways from the left inlet to the right outlet, as shown in the first image of Fig. 5(a). It can be seen that the velocity magnitude distributions are quite close for different cases (four images in the left column), because gas slippage in the porous medium with high porosity (or large pore radius) is insignificant and the Klinkenberg's effect can be neglected. As the porosity is reduced ($\varepsilon_{ps}$=0.5, middle column), the difference of |**u**| between different cases becomes obvious although they are qualitatively similar. A lower pressure leads to a higher apparent permeability of the permeable solid, resulting in stronger flow rate. When the porosity is further reduced ($\varepsilon_{ps}$=0.1, right column), the velocity magnitude difference becomes more remarkable. For the case of no Klinkenberg's effect, the preferred pathways for fluid flow are very different from those at higher porosity, as shown in the right image

of Fig. 5(a), due to the extremely low permeability in the permeable solid. However, the stronger the Klinkenberg's effect (from Fig. 5(b) to Fig. 5(d)), the closer the flow field is to that under high porosity. Normalized velocity magnitude $u_0=|\mathbf{u}|/|\mathbf{u}|_{max}$ is also calculated in the entire domain. Ten uniform ranges of $u_0$ are selected and the percent of the cells falling in each range is shown in Fig. 6. When porosity is high, the normalized velocity magnitude is almost the same; however it shows variations between different cases for low porosity, in consistent with above discussion related to Fig. 5. Fig.7 shows the global permeability for different cases. The permeability difference is obvious over the range of selected parameters. The various behaviors under different porosity and pressure are expected based on the above discussion.

In summary, without considering the Klinkenberg's effect, not only quantitatively but also qualitatively incorrect flow filed would be predicted. This incorrect prediction of flow, results in the misunderstanding of the transport mechanisms and erroneous values for permeability. In shale matrix where nanosize pores dominate [30], the apparent permeability will be much higher than the intrinsic permeability. Using the intrinsic permeability in the REV scale modeling will lead to underestimation of the fluid flow rate and consequently low estimation of the gas production curve [46, 47]. Finally, during the shale gas extraction process, the reservoir pressure is high initially, and then gradually decreases as the production proceeds. As shown in Fig. 7, the apparent permeability increases with the decrease of the pressure. Hence, a dynamic apparent permeability should be adopted in the reservoir simulations [48].

### C. Flow in porous media with fractures

In this section, the realistic porous structures of a shale matrix reproduced from the literature (see Fig. 13(a) in [42]) are considered, as shown in Fig. 8. Due to the intrapores between the grains of nonorganic matters [42], the black solid (nonorganic matters) in Fig. 8 now is considered to be permeable, but with a very low porosity of 0.05. The dark green solid (organic matter) is still permeable with a relatively higher porosity of 0.2. Volume fraction of the organic matter is 0.106. As shown in Fig. 8, the organic matter is embedded in the nonorganic matters; therefore, the transport of shale gas out of the organic matter will be extremely slow due to the quite low permeability of the black solid. Therefore, conductive pathways should be artificially generated. Currently, hydraulic fracturing is a method widely used to increase the permeability of a shale formation by extending and/or widening existing fractures and creating new ones through the injection of a pressurized fluid into shale reservoirs [49]. Here, we generate fractures in the reproduced shale matrix using a self-developed method called tree-like generation method. As shown in Fig. 8, the fracture system percolating the shale matrix consists of a main tree-branch at the center with several first and second sub-branches. With the fractures added, the volume fraction of the organic matter is slightly reduced to 0.097 and that of the fractures is 0.081. The size of the domain is 1164×536 lattices, with a resolution of 100nm. The typical aperture of the fractures is about 1μm. The generalized LB model for fluid flow through porous media with Klinkenberg's effect developed in Section III is applied to this system. The boundary conditions are the same as that in Section IV.B.

Fig. 9 shows distributions of the velocity magnitude in the shale matrix with and without fractures under pressure of 4000psi (Fig. 9(a)) and 100psi (Fig. 9(b)). Without fractures, fluid flow is stronger in the organic matter, which is expected due to the lower local flow resistance. With fractures added, the gas percolates the system through the fractures and the fluid flow is significantly enhanced. The Klinkenberg's effect becomes more important as the pressure reduces, leading to stronger fluid flow rate, as shown in Fig. 9. Fig. 10 displays the relationship between the global permeability and the pressure. For pressure of 4000psi, the global permeability of the system with fractures is about two orders of magnitude higher than that without fractures, indicating the significant importance of fractures for enhancing gas flow. The permeability difference becomes smaller as the pressure decreases. For pressure of 100psi, the global permeability for the system with fractures is $1.18\times10^{-14}$ m$^2$, only two times higher than that without fractures ($0.50\times10^{-14}$ m$^2$). This is because the gas mainly flows in the fractures, where Klinkenberg's effect is not obvious due to the relatively large fracture aperture of about 1μm. However, the Klinkenberg's effect is very strong in the shale matrix especially under low pressure. Thus, as the pressure decreases, the increase of the permeability for the system with fractures is not as significant as that without fractures, as shown in Fig. 10, thus reducing the difference.

## V. CONCLUSION

We proposed generalized Navier-Stokes equations for fluid flow through porous media that includes Klinkenberg's effect. Second-order Beskok and Karniadakis-Civan's correlation is employed to correct the apparent permeability based on the intrinsic permeability and Knudsen number. A LB model is developed for solving the generalized Navier-Stokes equations. Numerical simulations of fluid flow between two parallel plates filled with a porous medium have been carried out to validate the present model.

Flow in a porous medium with different components of varying porosity and permeability is studied. The simulation results show that the Klinkenberg's effect becomes stronger as *Kn* increases, either due to decrease of the porosity of the porous medium (This leads to the decrease of the characteristic pore radius) or the decease of the pressure (This leads to the increase of the mean free path). Global permeability of the porous medium increases as the Klinkenberg's effect becomes stronger. Without considering the Klinkenberg's effect, not only quantitatively but also qualitatively incorrect flow filed would be predicted. This incorrect prediction of flow, results in the misunderstanding of the transport mechanisms and erroneous values for global permeability. Flow in a porous medium with fractures is also investigated. It is found that with fractures added, the gas percolates the system through the fractures and the fluid flow is significantly enhanced.

The simulation results of the present study have a great implication to gas transport process in tight gas and shale gas reservoirs which pose a tremendous potential source for natural gas production. Distinguished characteristics of the shale matrix are that nanosize pores widely exist and the permeability is extremely low. Under such scenario, Klinkenberg's effect must be considered and apparent permeability should be adopted in the numerical modeling to predict the physically correct transport process. Using the intrinsic permeability will underestimate the gas flow rate and consequently generate inaccurate gas production curve. Besides, during the shale gas extraction process, the reservoir pressure dynamically changes, thus a dynamic apparent permeability should be adopted in the reservoir simulations based on our simulation results. The Klinkenberg's effect becomes increasingly important as the gas extraction proceeds because the pressure gradually decreases. Hydraulic fracturing facilitates the shale gas transport, and a small volume fraction of fractures generated in the shale matrix can significantly increase the global permeability.

# ACKNOWLEDGEMENT

The authors acknowledge the support of LANL's LDRD Program and Institutional Computing Program. The authors also thank the support of National Nature Science Foundation of China (51406145, 51320105004 and 51136004) and National Basic Research Program of China (973 Program, 2013CB228304).

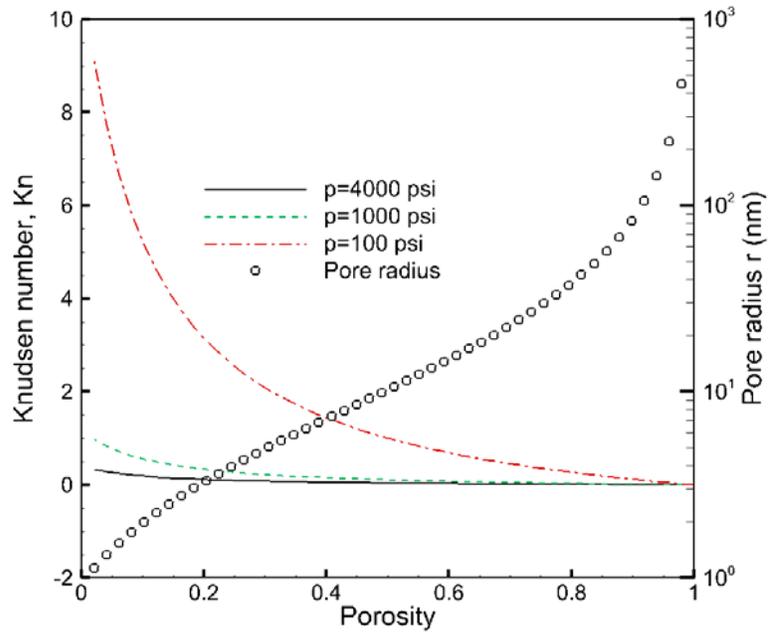

(a)

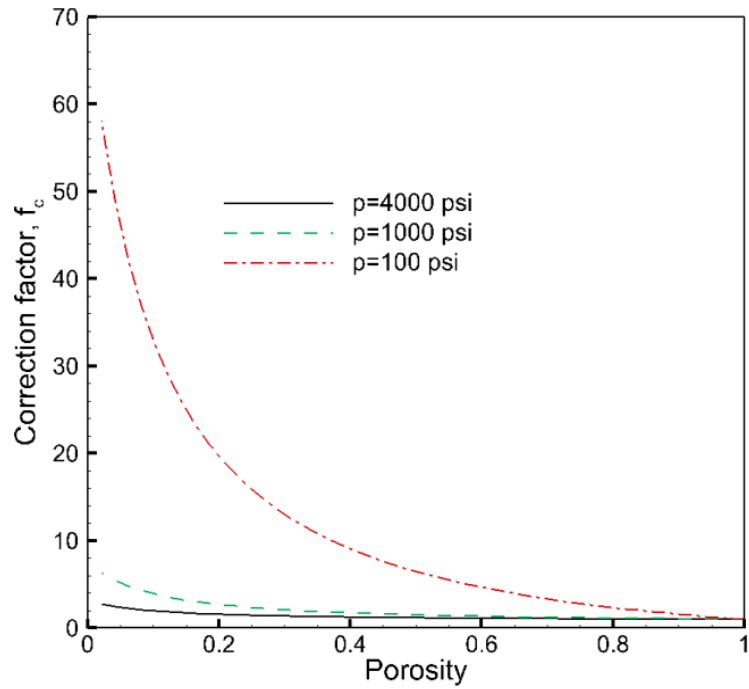

(b)

Fig. 1 (Color online) (a) Knudsen number, pore radius and (b) correction factor under different porosity

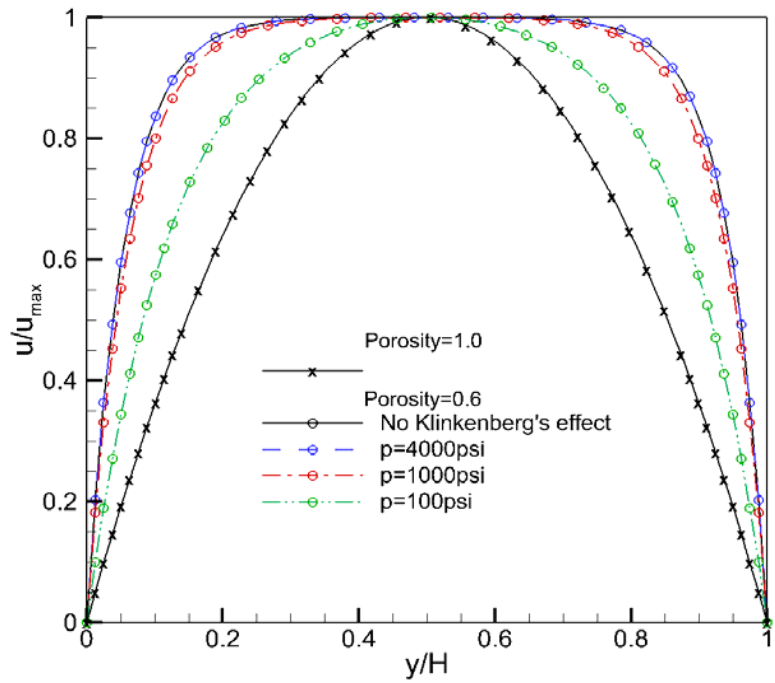

Fig. 2 (Color online) Velocity profiles between two plates filled with a porous medium. The simulation results (symbols) are in good agreement with the analytical solutions (lines).

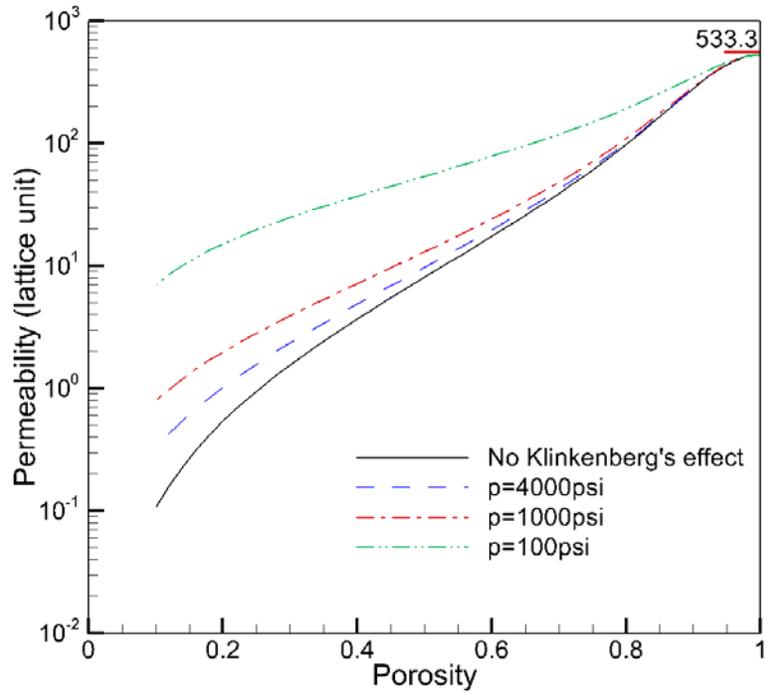

Fig. 3 (Color online) Relationship between the global permeability and porosity. The global permeability increases as the pressure decreases. The permeability difference between different cases increases as the porosity is reduced.

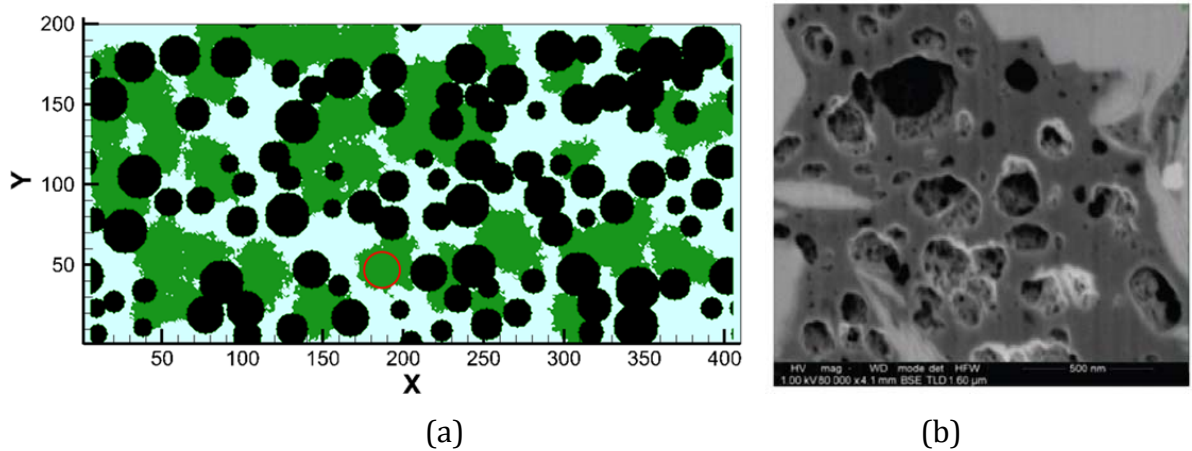

(a) (b)

Fig. 4 (Color online) (a) A heterogeneous porous medium with three components: pores (light blue), impermeable solid (dark circle) and permeable solid (dark green), which represents the pores, nonorganic matter and organic matter in shale matrix, respectively. The impermeable solid circle is generated by a self-developed overlapping tolerance circle method. The randomly distributed permeable solid is generated using the quartet structure generation set (QSGS) method [40]. Volume fraction of the impermeable solid and permeable solid is 0.4 and 0.3 respectively. (b) The nanoscale structures of the organic matter (reproduced from Fig. 9 in Ref. [33])

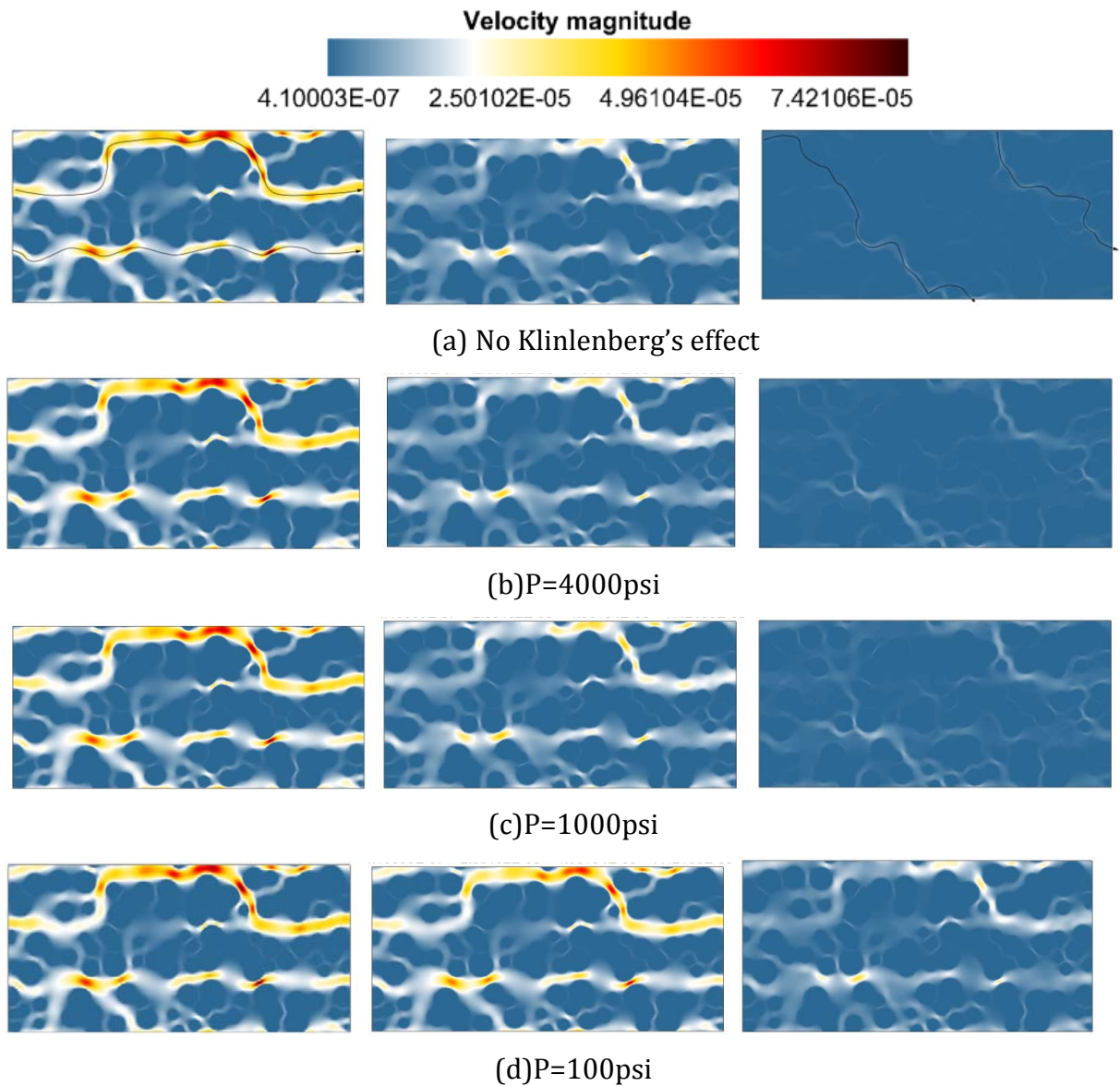

Fig. 5 Velocity magnitude distributions in the porous medium with and without Klinkenberg's effect. The porosity of the permeable solid in the left, middle and right column is 0.8, 0.5 and 0.1, respectively. As the Klinkenberg's effect becomes stronger with the decrease of the pressure, the flow rate increases due to the increased permeability in the permeable solids. The preferred pathways for fluid flow change (solid dark lines in Fig. 5(a)); therefore, without considering Klinkenberg's effect, not only quantitatively but also qualitatively incorrect flow filed would be predicted.

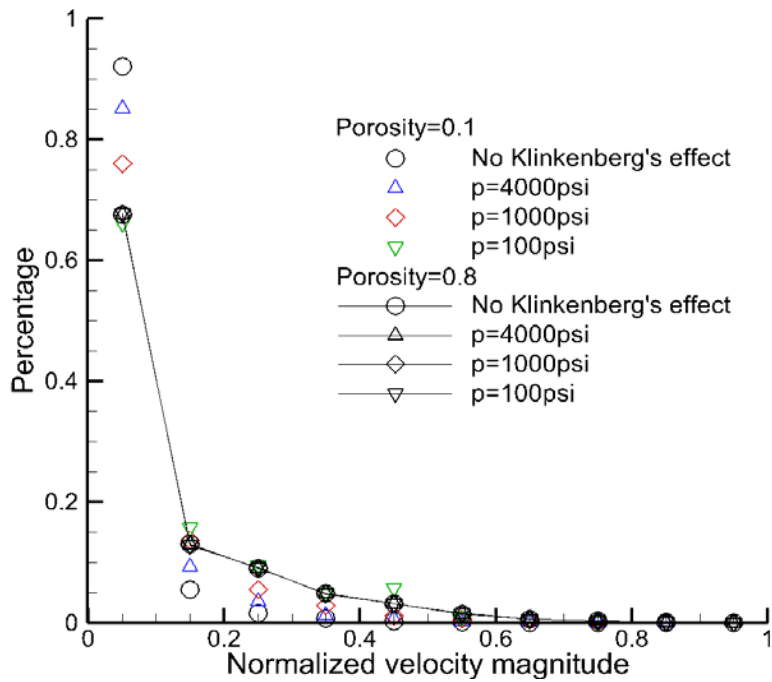

Fig. 6 Distributions of the normalized velocity magnitude. Normalized velocity magnitude $u_0=|\mathbf{u}|/|\mathbf{u}|_{max}$ are calculated in the entire domain. Ten uniform ranges of $u_0$ are selected and the percent of the cells falling in each range is calculated.

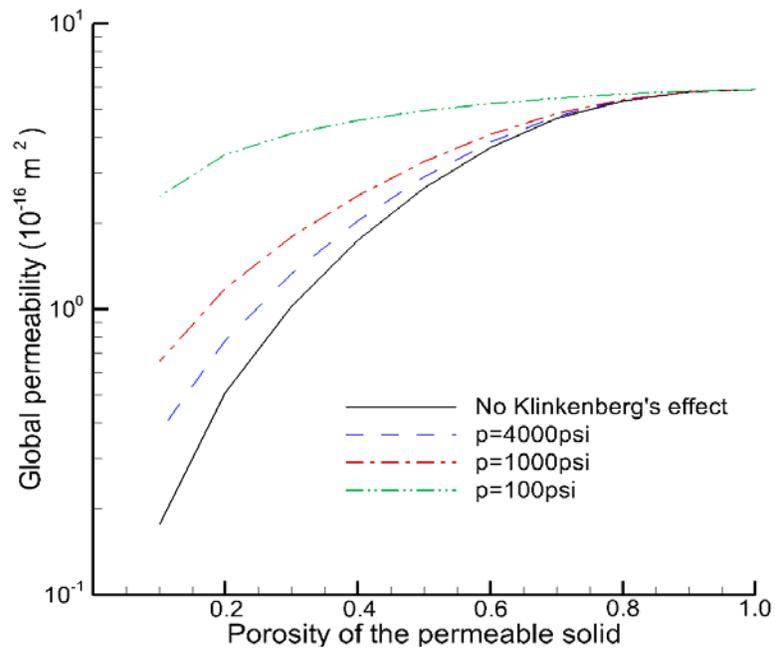

Fig. 7 (Color online) Relationship between the global permeability and porosity for cases with and without Klinkenberg's effect. When Klinkenberg's effect is considered, the global permeability increases as the pressure decreases.

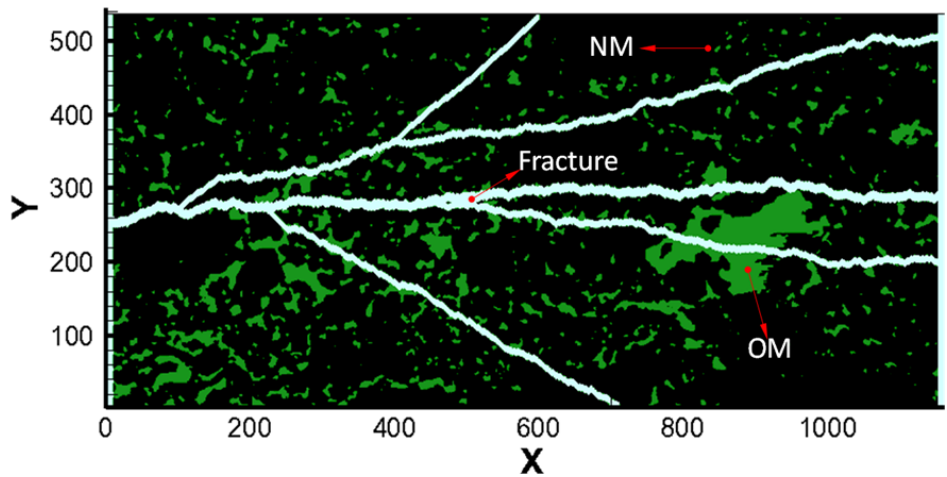

Fig. 8 Structures of a shale matrix with artificially generated fractures. The shale matrix is reproduced from Fig. 13 in [42]. The fractures (light blue) are generated using a self-developed method called tree-like generation method. The porosity of the organic matter (green) and nonorganic matter (black) is 0.05 and 0.2, respectively.

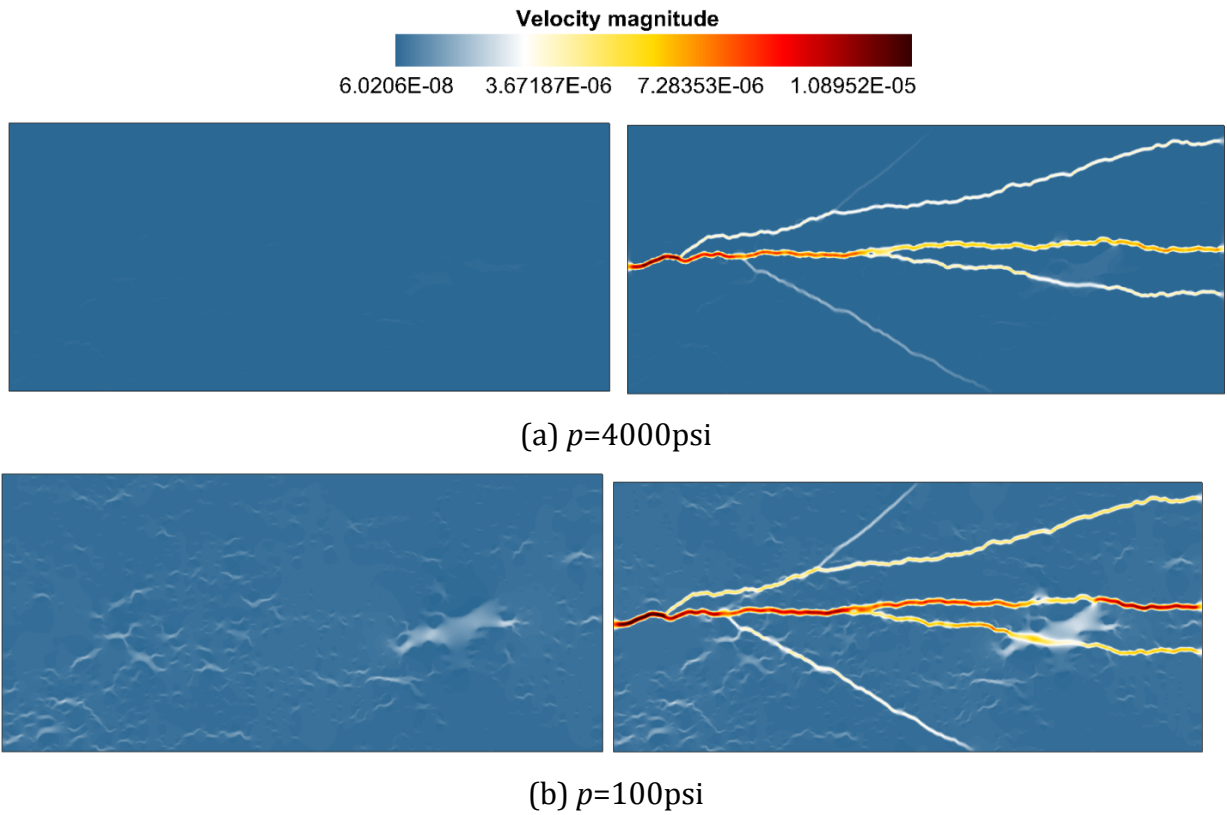

(a) *p*=4000psi

(b) *p*=100psi

Fig. 9 Distribution of velocity magnitude in shale matrix with (right) and without (left) fractures. With fractures, the velocity is greatly enhanced. The Klinkenberg's effect is obvious compared the velocity magnitude under different pressures.

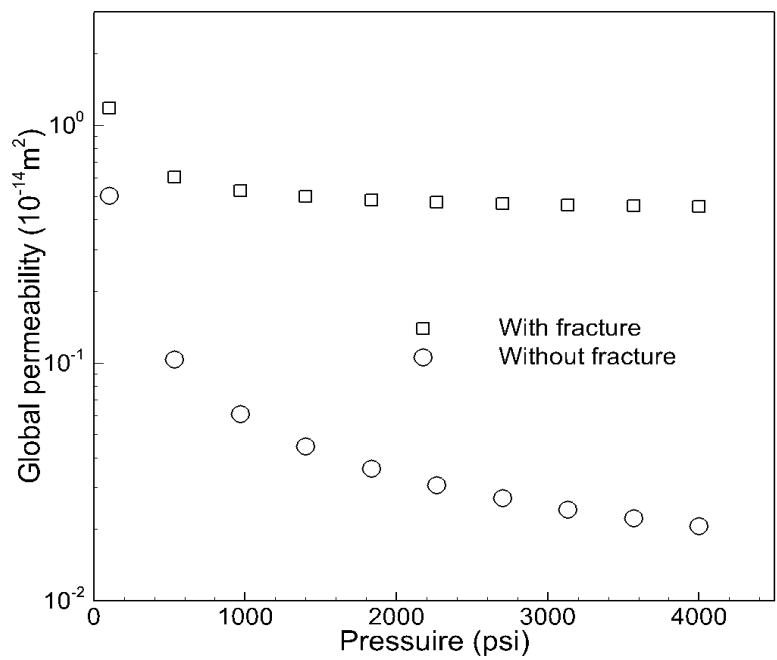

Fig. 10 The global permeability under different pressures